\documentstyle[12pt,aasms4]{article}

\begin{document}

\title{Models of OH Maser Variations in U Her}
\author{Stacy Palen, John D. Fix}
\affil{University of Iowa, Iowa City, IA, 52240}
\authoremail{palen@astro.washington.edu, jdf@astro.physics.uiowa.edu}

\begin{abstract}
Arecibo spectra of the mainline OH maser emission from U Her over more than a decade show variations of the OH emission over these time scales.  These observations are combined with high spatial resolution VLBA maps to investigate the changes in the velocities of the maser components.  Global properties of the dust shell, such as accelerations, variations in the pump and shell-wide magnetic field changes are examined as possibilities, and eliminated.  A solution to the problem involving plasma turbulence and the local magnetic field is introduced, and the relevant time scales of the turbulence are calculated.  The turbulent velocity field causes variations on time scales that are too long (of order centuries), while the turbulent magnetic field causes variations on appropriate time scales of a few years.  A line-of-sight model of the turbulence is developed and investigated.  The complete exploration of this solution requires extensive theoretical and observational work.  Possible avenues of investigation of the plasma turbulence model are presented.
\end{abstract}


\section{Introduction}
Radio observations of the OH maser emission from OH/IR stars have often been used to investigate properties of the dust shell, such as its shape \markcite{chapman1991}(Chapman, Cohen and Saikia 1991, \markcite{alcock1986} Alcock and Ross 1986), its density structure \markcite{maclow1996}(MacLow 1996), and its history \markcite{chapman1991}(Chapman, Cohen and Saikia 1991).   With a few exceptions, the maser emission from these stars has not been monitored over decade time scales, in part because the study of maser emission itself is only a few decades old.  U Orionis has been studied extensively by many groups since 1972, when an unusual flare in the 1612 emission occurred.  In particular, in 1991, Chapman, Cohen and Saikia reported results of a monitoring program of U Ori lasting 6 years.  Observing programs spanning decades should prove interesting, because the time scale for gas to cross the maser emitting shell is of order 10 years.  U Herculis is a particularly good candidate for maser studies, since it is relatively close (~385 pc \markcite{chapman1994}(Chapman, et al. 1994)) and has strong OH maser emission, with a peak maser flux that varies between 3 and 20 Jy in the 1665 and 1667 MHz lines. In 1977, Fix (1999) began a monitoring program of U Her and 11 other OH/IR stars at the Arecibo Radio Telescope.  His U Her data was indispensible to this project.

Section~\ref{observations} describes the Arecibo Radio Telescope (Arecibo) \footnote{The Arecibo Observatory is part of the National Astronomy and Ionospheric Center, which is operated by Cornell University under a cooperative agreement with the National Science Foundation.} and Very Long Baseline Array (VLBA) \footnote{The National Radio Astronomy Observatory is a facility of the National Science Foundation operated under cooperative agreement by Associated Universities, Inc.} observations of U Her, as well as the data reduction and analysis carried out prior to modeling.  Section 3 explains the need for and the construction of simulated sets of maser components.  We explore several explanations for the observed variations in Section~\ref{bulk}, and find that overall changes in the properties of the maser emitting shell do not fully explain the observed variations in the maser velocities.  A plasma turbulence model is presented in Section~\ref{plasmaturb}, and shown to be a promising line of inquiry.  Section~\ref{discussion} contains a summary and discussion of these results, as well as suggestions for future work.
\section{Observations, Analysis, and Interpretations}

\label{observations}

\subsection{Observations}

Observations of U Her were obtained in both maser main lines (1665 and 1667 MHz), and both right circular polarization (RCP) and left circular polarization (LCP).  The total bandpass of all observations was 25 km/s ($\approx$ 488 kHz), although the spectral resolution varied with instrument as described below.

Spectra of the main line maser emission from U Her were obtained at Arecibo at four epochs: 1977.866 (henceforth referred to as 1977), 1979.151 (1979a), 1979.953 (1979b), and 1992.384 (1992).  The spectral resolution at most epochs was 0.05 km/s.   At 1667 MHz, the 1979a data were of insufficient spectral resolution, and so are not included in the full discussion, but reserved until section 4.1.  

On 1995.512 (1995), we used the VLBA, supplemented by one antenna of the Very Large Array (VLA) to map the OH emission.  Approximately 3 hours of data were obtained on the target source. The data were correlated using the VLBA correlator in Socorro, NM.  The spectral resolution was 0.1 km/s.  The beam size was approximately 10$\times$4 mas.  3C273 was used as a flux calibrator, and 3C345 was chosen as a phase calibrator.  The \textit{a priori} calibration of these VLBA data was carried out using the spectral-line calibration methods described in the AIPS Cookbook in Chapters 4, 8 and 9.  Remaining errors were removed using a loop of self-calibration and CLEAN algorithms.  

The red-shifted portion of the 1667 MHz line (from the back of the shell) was too faint to be mapped with the VLBA.  Also in 1667 MHz, the VLA antenna recorded interference.  The consequent loss of short telescope spacings reduced the sensitivity of the 1667 MHz maps to diffuse emission at angular scales longer than 40 mas.  Spectra were created from the VLBA single antenna data.  These spectra were significantly noisier than the Arecibo data, since the sensitivity of the single antenna can not compare to Arecibo.  This is evident in Figure~\ref{spectra}.

\subsection{Analysis}

Each spectrum was deconstructed into a series of Gaussian components using a least-squares fitting routine.  Between 19 and 30 components were required to adequately characterize each spectrum. 

The overall maser emitting region was resolved into a number of small maser components as shown in Figure~\ref{maps}.  This ``spotty'' appearance was evident in all the maps, in both polarizations, and both main lines.  The AIPS task SAD was used to find the maser components in the individual maps.  The integrated amplitudes of these maser components were then grouped by location across the channels to create a spectrum for each component.  This resulted in a set of spatially resolved components (``spatial components'') each with an associated spectrum.

\subsection{Interpretations}
Figure~\ref{spectra} shows the dramatic changes in the Arecibo spectra between 1977 and 1992.  The changes in the spectra can not be described as simple amplitude variations of the entire line, as would be produced in response to variations in the intensity of the central star. 

Only 7 of the 80 maser spots in the maps were spatially coincident, indicating that the individual maser spots are highly polarized, possibly via the Cook mechanism (Cook 1975).  The Cook mechanism balances velocity gradients and magnetic field gradients so that only one polarization is amplified.  In contrast, the spectra are not highly polarized, implying that an annular region of approximately constant projected outflow velocity may have several components.  These two observations together indicate that it is likely that an apparent Zeeman pair in a spectrum is actually a pair of spatially distinct Cook components.

The spectra of the spatial components were compared with the Gaussian components derived directly from fitting Gaussian components to the VLBA spectra.  In the case of the 1665 MHz data, 87\% of the Gaussian components could be identified with a particular spatial component.  8\% of the Gaussian components could not be identified in the maps.  The amplitudes of these components were small.  5\% had more than one spectral component that could be identified with one spatial location.  For the 1667 MHz observations, only 54\% of the Gaussian components could be identified with the spatial components.  This is probably due to the decreased flux in 1667 MHz, and the decreased sensitivity at larger size scales.  These results imply that a component by component analysis of the Arecibo spectra yields variations of real, physically distinct components.

\section{Simulated Data}
In order to test whether model results are meaningful, 25 sets of simulated components were created.  Each set of simulated components consists of a number of velocities and associated errors.  Note that a spectrum was \textit{not} produced by this process.  The components have neither width nor amplitude, only velocity.  This is the primary parameter of interest in the models below.

All of the relevant information was created using the pseudo-random number generator distributed with ANSI-C.  The generator was seeded with a number between 1 and 100.  The generator creates a long list of random numbers, from which we created velocities, errors, and numbers of components.

The first step was to choose the number of components in a data set.  Since the real data sets contained between 19 and 30 components each, the first 25 random numbers were transformed to values between 19 and 30.  

The velocities of the components in a data set were found by transforming the next 25 (say) random numbers into velocities ranging between \nobreak{-4} and \nobreak{-21}, to simulate the range of velocities found in the real spectra.  Each simulated data set was created from different lists of random numbers.

We also created ``errors'' in the simulated components.  The errors in the real components ranged between 0.001 and 0.02 km/s, depending roughly on the amplitude of the component, but independent of the velocity.  The ``errors'' in the simulated components were created by transforming the random numbers to numbers between 0.001 and 0.02 km/s.  The lists of simulated velocities and simulated errors were collated so that for each simulated set of components, there were between 19 and 30 velocities and associated errors.

\section{Models Involving Overall Properties of the Shell}
\label{bulk}
There are at least three ways in which the maser components may vary due to changes in the overall properties of the maser shell.  The first is that the masers may be moving outward at constant velocity, through regions of varying pump intensity.  In this case, components will remain at the same velocity, while changing in amplitude.  The second possibility is an acceleration of the shell, either radial or rotational, which changes the velocities of the maser components.  The third possibility is a change in the overall magnetic field, which can be observed in changes in Zeeman split components, in which the members of an LCP/RCP pair will change velocity in opposite directions, either moving closer together along the velocity axis, or farther apart. 

All of the models involving bulk properties of the shell were investigated using the following algorithm.  An initial epoch was chosen, then the model was propagated forward in time, modifying each component of the initial epoch accordingly.  For example, if the model predicts that between 1977 and 1979, the components will shift redward in velocity by 0.25 km/s, then 0.25 was subtracted from the velocity of each 1977 component.  These modified components were then compared to the 1979 components to look for matches.  A match was found when the difference in the predicted and actual velocities was less than or equal to twice the sum of their error bars.  

In order to understand the meaning of the statistics, the model was also applied to pairs of simulated data, to compare with the number of matches expected purely by chance.  In order for a model to be considered successful, it must accurately predict the changes in the central velocities of the components from one epoch to the next.  The percentage of components matched in the real data must be significantly larger than the percentage of components matched in the simulated data.  In general, there were 24 comparisons of actual components (2 frequencies, 2 polarizations, and 4 epochs, combined pairwise by epoch), and 51 comparisons of simulated components.

The parameter space for each of the models was completely explored.  Each parameter in each model was varied from a theoretical lower bound to a theoretical upper bound, in a series of incremental steps.  For each value of these parameters, the number of matches between epochs and between simulated component sets was recalculated.  The step size was always carefully chosen so that the model predictions would be adequately sampled.  This is explained further for each model below.  The number of parameters in each model was small enough that this method of investigation was not overly costly in terms of CPU time. 

\subsection{Changes in the Pump}
If the masing components move outward at constant velocity, then each individual component remains at the same velocity (projected along the line of sight) while increasing or decreasing in amplitude.  Spectral components with have the same velocity in more than one epoch.  The results of this model are shown in Figure~\ref{histogram}, and expressed as a percentage of the total number of matching components that might have been found.  These two binned distributions yield a reduced chi-square of 1.4, indicating that the distributions are statistically identical.  This model fails to fit real data better than it fits simulated data, implying that the matches in the real data are consistent with chance.

\subsection{Radial Acceleration}
A radial acceleration changes the velocity of the material as it moves out through the shell.  Because of projection effects, the radial acceleration will be most apparent at the front and the rear of the shell, where the expansion is most nearly along the line of sight.  A radial acceleration at the limb of the shell will not cause an observable change in the velocity of a maser component located there.  The emission from the front and back of the shell is located at the outside of the line profile, and so we expect to see the greatest changes due to radial acceleration in these regions of the line.

The change in the projected velocity due to a radial acceleration is given by
\begin{equation}
\Delta v_{pr}=\Delta v_{exp} \cos\theta
\end{equation}
where $\Delta v_{exp}$ is the total change in the expansion velocity due to acceleration, and $\theta$ is the angle between the maser and the projected center of the shell. The cosine of the angle is equal to the ratio of the observed velocity to the expansion velocity, $v_{obs}/v_{exp}$, so that equation (1) becomes 
\begin{equation}
\Delta v_{pr}=\Delta v_{exp}\frac{v_{obs}}{v_{exp}}
\end{equation}
The component will be shifted towards the outside of the line for a positive radial acceleration, and towards the center of the line for a negative acceleration.  Red-shifted and blue-shifted lines will move in opposite directions, but both will move towards or away from the center if they are affected by the same acceleration.

The projected velocity at a later epoch, $v_{r,b}$, is related to the projected velocity at an earlier epoch, $V_{obs}$ by
\begin{equation}
v_{r,b}=v_{obs}(1\pm\frac{\Delta v_{exp}}{v_{exp}})
\end{equation}
where $v_{r,b}$ indicates the red- or blue-shifted velocity, and the direction of the acceleration is given by the sign of $\Delta v_{exp}$.

From the largest and smallest velocities of the maser components of U Her, the expansion velocity is between 6 and 8 km/s.  This expansion velocity is expected to remain approximately constant since the dust has already condensed (Fix and Cobb 1987).  With this range of values for the expansion velocity, the theoretically expected values for $\Delta v_{exp}/v_{exp}$ are less than 0.2 since the expansion velocity changes by at most a few tenths of a km/s per decade.  Varying the range of $\Delta v_{exp}/v_{exp}$ between -2 and 2 easily covers the possible range of values, and is in fact much larger than necessary, as it allows even the innermost components to be completely shifted out of the line profile.  A step size of 0.0005 was used.  The errors in the centroids of the components were greater than 0.001.  This step size is approximately 1/8 of twice the sum of the error bars for a pair of components.  This is the criterion for finding components at the same velocity, and so this step size should adequately sample the radial acceleration model.

The average percentage of components which could be identified between epochs was 39\% for the real data, and 35\% for the simulated data.  The reduced chi-square of the two distributions (real and simulated data) of these percentages is 1.9, consistent with identical distributions for these small number statistics.  

One of the assumptions of this model is that the radial acceleration is spherically symmetric (although the shell need not be).  A consequence of this assumption is that the front and the rear of the shell undergo the same acceleration.  This constraint was relaxed by allowing the front and back of the shell to have different accelerations.  In this case, the reduced chi-square of the distributions of percentages of matched components for real and simulated data was found to be 0.46.  This is still consistent with all of the matches in the actual data being found purely by chance.  

A second consequence of this assumption is that many components undergo the same acceleration. If this is not the case, this model does not apply, and we must try to model the components individually, and look for changes in the velocities which are linear in time (see section 4.5).

\subsection{Magnetic Field}
The maser emission from OH/IR stars is often polarized.  This indicates that magnetic fields are present which are strong enough to Zeeman split the OH lines.  In order to investigate changes in the global magnetic field, we must first look for Zeeman pairs.  

The VLBA maps provide the most compelling evidence of magnetic fields strong enough to Zeeman split the lines.  In the 1665 MHz emission, only 7 of 47 maser spots appeared in the same location in both RCP and LCP emission.  In 1667 MHz, none of the 33 maser components appeared in both polarizations.  This implies that at least 80\% of the components are polarized, possibly via the Cook mechanism \markcite{cook1975} (Cook 1975).  As mentioned above, this implies that we are unlikely to find a pair of Zeeman components in the spectra which are real.  The probability that an apparent Zeeman pair is actually a pair of spatially distinct Cook components is high.      

The importance of the Cook mechanism can be verified directly from the spectral data.  The main-line maser emission is thought to arise in a narrow region of the shell (see for example, Collison and Nedoluha, 1994).  If the magnetic field is constant (or approximately so) throughout the region, then the Zeeman splitting will also remain constant throughout the shell.  In this case, the Cook mechanism may be ignored, since there is no magnetic field gradient.  There should be a constant splitting of each of the LCP/RCP pairs, since each component experiences the same magnetic field.  If the magnetic field gradient is important, few pairs of LCP and RCP components will be found with the same Zeeman splitting.

The number of Zeeman pairs was calculated by choosing a component in one polarization, then searching in the other polarization for a component offset by a prescribed Zeeman splitting.  The range of the splitting was -2 to 2 km/s, and the step size was 0.001 km/s.  These parameters were chosen to encompass the entire range of possible values, and to provide a step size smaller than the errors in the velocities of the components.  The average percentage of matched components for the actual data was 34.5\%, while the average percentage of matched components for the simulated data was 32.5\%.  When the 1995 data, with its larger spectral resolution, was removed from consideration, the average percentage of matched components in the actual data dropped to 27\%.  Both of these results are consistent with the results from the simulated data.

\subsection{Linear and Quadratic Changes in Time}
The investigations described above lead us to conclude that we cannot model the changes in the velocities of the components as global changes caused by constant velocity motions, radial or rotational accelerations, or global magnetic fields.  However, it may be possible that all components, while subject to varying physical conditions, are still constrained to change in the same way over time.  For example, all components could be subject to an acceleration that varies between one component and another.  Models of this type will not give any detailed information about the mechanism of the change, but they will tell us whether the same mechanism operates on each maser feature throughout the observed time period.  We note that if the variations are caused by a combination of the above scenarios, then this model should be able to characterize those variations.

The number of 1977 components was in some cases (e.g. 1665 RCP) greater than the number of components in the 1979b or 1992 data.  There may be redundancies in the alignment of components, so that two or more 1977 components may be aligned with one of the 1979b or 1992 components.  Since it is not possible to determine which of these is the ``correct'' pairing, we simply counted the number of 1977 data points for which a match may be made.  Because of this, the number of successful projections was occasionally higher than the smallest number of components in the three or four epochs.  

The four epoch case was completely consistent with the simulated data, implying that the matched components are likely to have been found by chance.  The three epoch case is less tightly constrained, and more matching components were found.  However, the largest percentage of actual matched components was no more than 5\% higher than the highest percentage of simulated matched components.  These results are not inconsistent with the results expected purely by chance.

For completeness, we investigated an acceleration which is variable in time.  An acceleration which changes over time will produce a velocity that varies quadratically in time.  For example, if the radiation pressure changes, the force on the gas will change, and the acceleration of the gas particles will change.  The possibility of a quadratic variation of the velocity was investigated, and it was found that the number of free parameters in this model is so large that a projected component can be found in the third or fourth epoch nearly 100\% of the time for both real and simulated data.

\section{Plasma Turbulence}
\label{plasmaturb}
We have now shown that the changes in the velocity and polarization structure of the OH main line maser emission from U Her over decade time scales can not be completely explained by overall properties of the shell, such as accelerations, or global magnetic field changes.  We are driven to consider explanations which do not involve properties of the shell as a whole, but rather can affect each maser component differently.  We know that the magnetic field must be important in these masers because nearly all of the spatial and spectral components are polarized.  The masers arise in regions with charge-carrying dust grains moving at a drift velocity which is of order tens of km/s \markcite{collison1992} (Collison and Fix 1992).  This is about 100 times the thermal speed of the gas derived from the full widths of the spectral components.  There is no direct evidence that the dust grains carry charge, although it has long been assumed that they do.  It seems plausible that the grains are collisionally charged by interacting with the ionized gas and its electrons, and theoretical models including charged dust match well with infrared observations \markcite{zubko1998} (Zubko 1998).  Plasma turbulence may arise, perhaps via an instability such as those described in plasma kinetic theory.  We investigate whether this turbulence could produce changes on the observed time scales.

\subsection{Time Scales of Turbulence}
In the data considered so far, the shortest time scale investigated was a bit more than two years.  A better constraint on the time scales may be found by comparing observations closer in time.  A second set of observations of U Her were taken on 1979.151 (1979a), about ten months before the 1979.953 (1979b) data set used to investigate the overall variations of the shell.  Only the 1665 MHz observations were of high enough spectral resolution to compare with the rest of the data, and so the 1979a observations were not included in the larger study.  The 1979a data was subjected to the models described in section 4.  These models tended to fit the 1979a and 1979b combination of epochs better than any other pair of epochs, but still, the percentage of components which could be identified across epochs was consistent with, or only marginally better (1-2\%) than in the simulated components.  Interestingly, even the constant velocity model failed to fit these epochs which were separated by less than one year.  This implies that the time scales of the variations of the individual components may be of order months.

The time scale for a turbulent eddy of size L to rearrange is 
\begin{equation}
\tau=\frac{L}{\delta v_t}
\end{equation}
where $\delta v_t$ is the ``turbulence velocity'', which is usually less than or of the order of the thermal velocity, or the Alfv\'{e}n speed.  In the case of U Her, we can calculate an upper bound to this length scale since the masers are partially resolved.  From the period-luminosity relation for Mira variables, the distance to U Her is estimated to be 385 pc \markcite{chapman1994} (Chapman et al 1994).  Feast (1989) quotes an error in the derived magnitude using this method of 0.14.  This yields an error in the distance of 26 pc (7\%).  From the maps, the average angular size of the maser emission regions is about 20 mas.  This gives a linear size of $\approx 1 \times 10^{14}$ cm for the regions which produce the maser emission.  If the turbulence velocity is the thermal velocity of the gas ($\approx 0.2$km/s=$2 \times 10^{4}$cm/s), then the time scale for complete change of the turbulent medium is about 200 years.  This is much longer than the time scale of the variations.  The simple turbulent motions of the gas cannot explain variations of the maser components on the time scales observed.

Suppose instead that the magnetic field variations drive the appearance and disappearance of the maser spots.  This possibility is supported by the observed importance of the Cook mechanism.  If the magnetic field is tied to the charged dust particles, then turbulent features will cross the maser emitting region at the dust drift velocity, given by Collison and Fix (1992) as $\approx20$ km/s.  This leads to a time scale of a bit less than two years.  While it is not certain whether the dust grains and the magnetic fields travel together, this seems like a reasonable assumption, since the $\approx 1$ mGauss fields inferred from the Zeeman splitting cannot be generated at the central star, and thus must be generated {\it in situ} by the motion of the charged particles.  Note that the magnetic field gradient is the important quantity in the Cook mechanism, not the magnitude of the magnetic field, and so this calculation is an upper bound on the time scale, since the turbulent magnetic field may have large enough gradients for the Cook mechanism to be rendered ineffective on length and time scales much smaller than the ones calculated here.

\subsection{Plasma Turbulence Along the Line of Sight: Theory and Model Description}

One way to investigate the validity of a plasma turbulence model is to investigate whether plasma turbulence can cause bright, polarized components with the same probability as that inferred from the filling factor of the masers in the radiation shell.  For example, if masers are produced in 20\% of the shell, then we would like to know if plasma turbulence produces bright, polarized components in 20\% of simulated lines of sight.

The filling factor of the observed maser emitting regions was found by summing the areas of the maser components, and dividing by the area of the shell projected on the sky.  The projected shell was assumed circular, with a diameter of $\approx 0.5$ arcseconds, as indicated by the distribution of the maser emission in the maps.  For the 1665 MHz maps, the filling factor of all the components at all velocities is 0.11.  For the 1667 MHz maps, the filling factor is 0.05.  However, since the back portion of the shell was too faint to map in 1667 MHz emission, the filling factor for the 1667 MHz emission may be as much as 0.1.  The filling factor to which the following model results were compared was 0.1.

We can reduce the plasma turbulence problem to one dimension by considering the maser as a line integral along the line of sight.  In each incremental path length $ds$, a given velocity $v$, and magnetic field $b$ are present.  These values of $v$ and $b$ are the small-scale turbulent velocities and magnetic fields, not the large-scale expansion velocity, or overall magnetic field.  We can calculate an effective velocity of emission from this incremental path length from the Doppler-shifted frequencies and the Zeeman splitting of the line due to the magnetic field 
\begin{equation}
v_{eff}=v \pm \frac{\gamma c}{f}B  \label{eq:effvel}
\end{equation}
where $c$ is the speed of light, $f$ is the frequency in MHz of the transition at rest (1665 or 1667), and $\gamma$ is the frequency splitting due to the Zeeman effect.  The sign indicates RCP or LCP shifts.  In order for amplification to occur, we must have a large number of path lengths $ds$ with the same effective velocity.  Collecting the effective velocities of the path lengths into bins and making a histogram shows the velocity coherence along the line of sight.  

In general, each path length will not be saturated.  If one of the bins contains more than the number of path lengths required for the maser to become saturated, then the maser will be amplified.  A component is considered bright and polarized if the ratio of the maximum bin in RCP to the maximum bin in LCP is greater than 1.3 or less than 0.7.  This criterion means simply that the maser component must be at least 30\% polarized in order to resemble a Cook polarized component.  

MHD turbulence theory shows that the power spectrum of turbulent velocities and magnetic fields is a power-law.  In a one dimensional case, the power spectrum is given by
\begin{equation}
P(k)=\tilde{A}^2(k)\propto k^{-a},
\end{equation}
where $\tilde{A}$ is the Fourier transform of a function in velocity space (e.g. the velocity, $v(x)$), and $k$ is the wave number in the Fourier transform space.  In the case of Kolmogorov turbulence, the exponent indicating the steepness of the power law, $a$, is 5/3.  We adopt this special case as a first approximation of the problem.  In our case, however, we may have turbulence on size scales larger than the size of the emitting region, $l_{0}$.  These will enter into the power spectrum as a flattening of the slope to a constant near $k$=0.  The point of turnover is given by $k_{0}=2\pi/l_{0}$.  We can express this modified Kolmogorov power spectrum in functional form as 
\begin{equation}
P(k) \propto{\left( 1+ \left( \frac{k}{k_0} \right)^2 \right)^{-5/6}}.
\end{equation}

In order to generate the velocity and magnetic field distribution, we must randomize the power spectrum given by MHD turbulence theory.  This guarantees that the general behavior of the velocities and magnetic fields agrees with what is expected from theory.  Typically, the power spectrum is randomized by multiplying by a random, complex number
\begin{equation}
\tilde{A} (k) = \sqrt{\frac{P(k)}{2}}(D+iF).
\end{equation}
$D$ and $F$ are both zero mean, Gaussian distributed, unit standard deviation numbers \markcite{spangler1998}(Spangler 1998).  This equation is additionally constrained by the requirement that $A(x)$ (the velocity or magnetic field vector) is real.  This requirement will be met when $\tilde{A}^{*}(k)=\tilde{A}(-k)$.  By even-odd symmetry, this condition constrains $F$ to be 0 when $\tilde{A}(k)$ is the modified Kolmogorov spectrum being considered here.  $A(x)$ is the inverse Fourier transform of $\tilde{A}(k)$
\begin{equation}
A(x)=\int_{-\infty}^{\infty} dk \exp (-2 \pi i k \cdot x) \tilde{A}(k).
\end{equation}
The velocity is given by $A(x)$.  
	The magnetic field, $b(x)$, was calculated as a linear combination of the velocity, $v(x)$, and a statistically independent function $b_{i}(x)$.  The function $b_{i}(x)$ was calculated in the same manner as $v(x)$, using a different amplitude, and a different set of random numbers $D$ in the randomization of the power spectrum.  The magnetic field is then given by 
\begin{equation}
b(x)=\alpha v(x)+(1-\alpha)b_{i}(x),
\end{equation}
where $\alpha$ is an adjustable parameter, with values between 0 and 1, which indicates the degree of correlation between the magnetic and velocity fields.  Plots of the randomized quantities $v(x)$ and $b(x)$ are shown in Figure~\ref{fields}.  

Once the velocity and magnetic fields were calculated, the effective velocity, Equation~\ref{eq:effvel}, was calculated, and path lengths grouped in bins by this value.  The bin width was chosen to be the spectral resolution of the Arecibo observations (0.05 km/s).  The resulting distribution was compared to the criterion for a bright, polarized component.  The model was recalculated for many different sets of the random variable $D$, which is equivalent to calculating the model along many lines of sight.  The probability of the production of a bright, polarized component was calculated, and compared to the filling factor of the maser components.  This probability was calculated for various values of the free parameters $k_{0}$, $\alpha$, $C_{vel}$, and $C_{mag}$.  

A subset of the results of this model are shown in Figure~\ref{turbmod}.  The strongest dependence in the model was on the correlation parameter $\alpha$.  Cases in which the magnetic and velocity fields were very highly correlated yielded the correct filling factor in only 11\% of trials.  Generally, the highly correlated case led to too many components, or none, so that either 20-40\% of the shell was filled, or there was no maser action.  A partial correlation ($\alpha = 0.5$) also gave the correct filling factor in only 11\% of trials.  For a nearly uncorrelated case, ($\alpha=0.01$), the correct filling factor was obtained 1/3 of the time.  This implies that the proper filling factor can be achieved through plasma turbulence, and that the velocity and magnetic fields are most likely uncorrelated.  None of the other parameters showed a trend as clear as the dependence on $\alpha$.  Overall, 19\% of the values of the parameters yielded filling factors that matched the observations.

\section{Discussion}
\label{discussion}
In this study, we have investigated three possible global explanations of the variations of the OH maser emission from U Her.  The first possibility was a movement of the maser, at constant velocity, through regions where the pump altered in strength.  The second possibility was an acceleration, either radial or rotational, of the shell.  The third was a change in the global magnetic field which causes the polarization of the maser components.  All three of these possibilities were investigated as bulk properties of the shell, and as properties which changed within the shell, and were peculiar to each maser.  None of these possibilities completely characterized the behavior of the main-line maser emission from U Her.  

Since the variations cannot be described by overall properties of the shell, and cannot even be modeled to vary in the same manner over time everywhere in the shell, we were driven to seek other explanations.  One promising explanation is that plasma turbulence effects alter the magnetic field in the emission region enough to remove the amplification of the masers by the Cook mechanism.  If the turbulent magnetic fields are carried by the dust grains, the plasma turbulence model produces changes on time scales of less than one year, which is in agreement with observed time scales of variation.  Also, it is possible to create shells with the correct filling factors using this model.

This is a radical departure from the usual school of thought for at least two reasons.  The first is that the determining factor in maser emission seems to be the magnetic field gradient, rather than the velocity gradient along the line of sight, since it is the turbulent magnetic field which varies on the observed time scales.  The second is that the coherent turbulent ``bits'' along the line of sight which add up to an amplified, polarized maser component are not necessarily physically contiguous.  There may be larger regions of incoherent plasma between the coherent regions, and still the masers will be amplified and polarized.  Thus, these masers can not be thought of as entities, or even preferred lines of sight, since the whole line of sight through the dust shell is not necessarily involved in the amplification.

\acknowledgments

\newpage
\figcaption[fig1a.ps 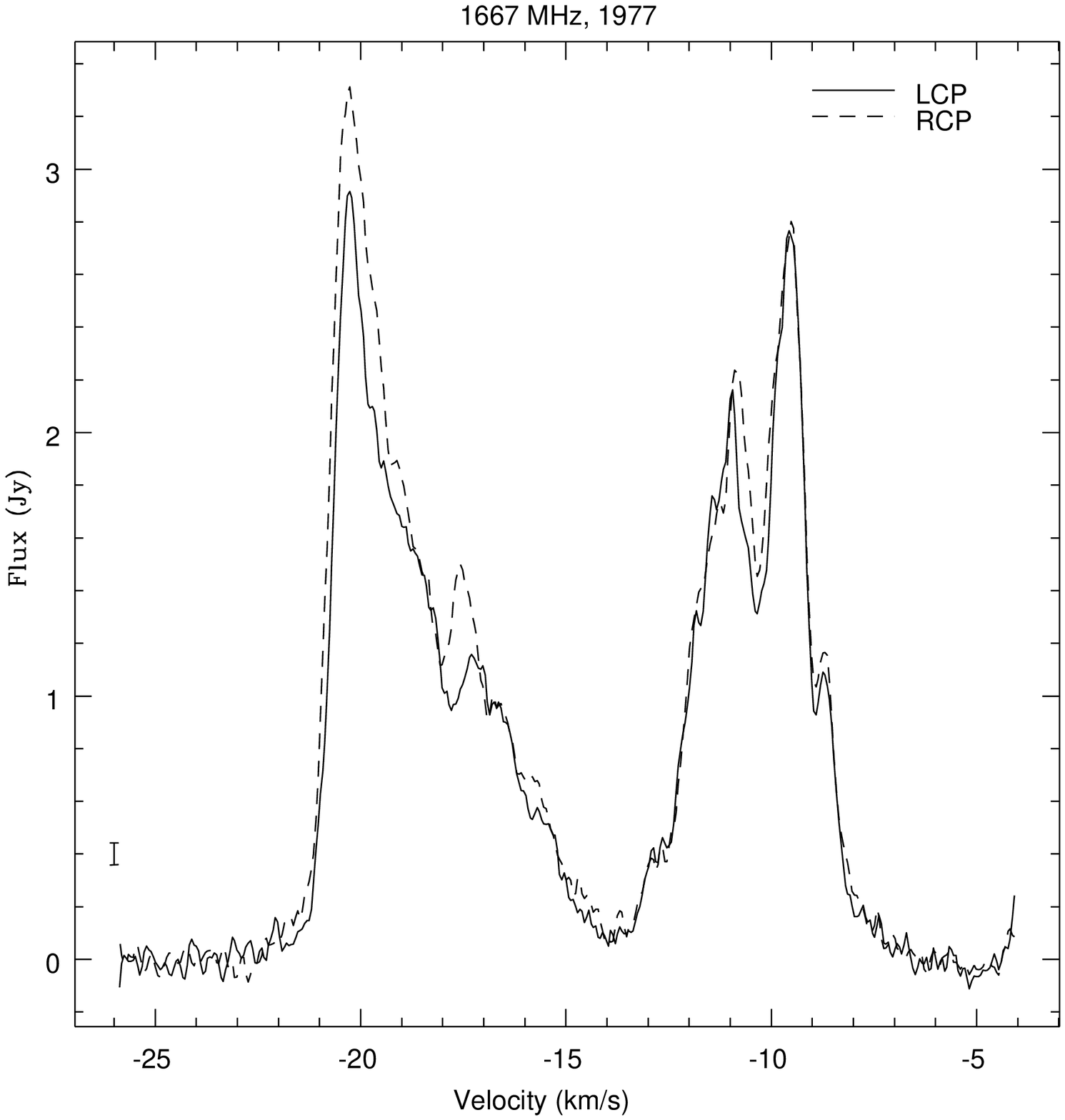 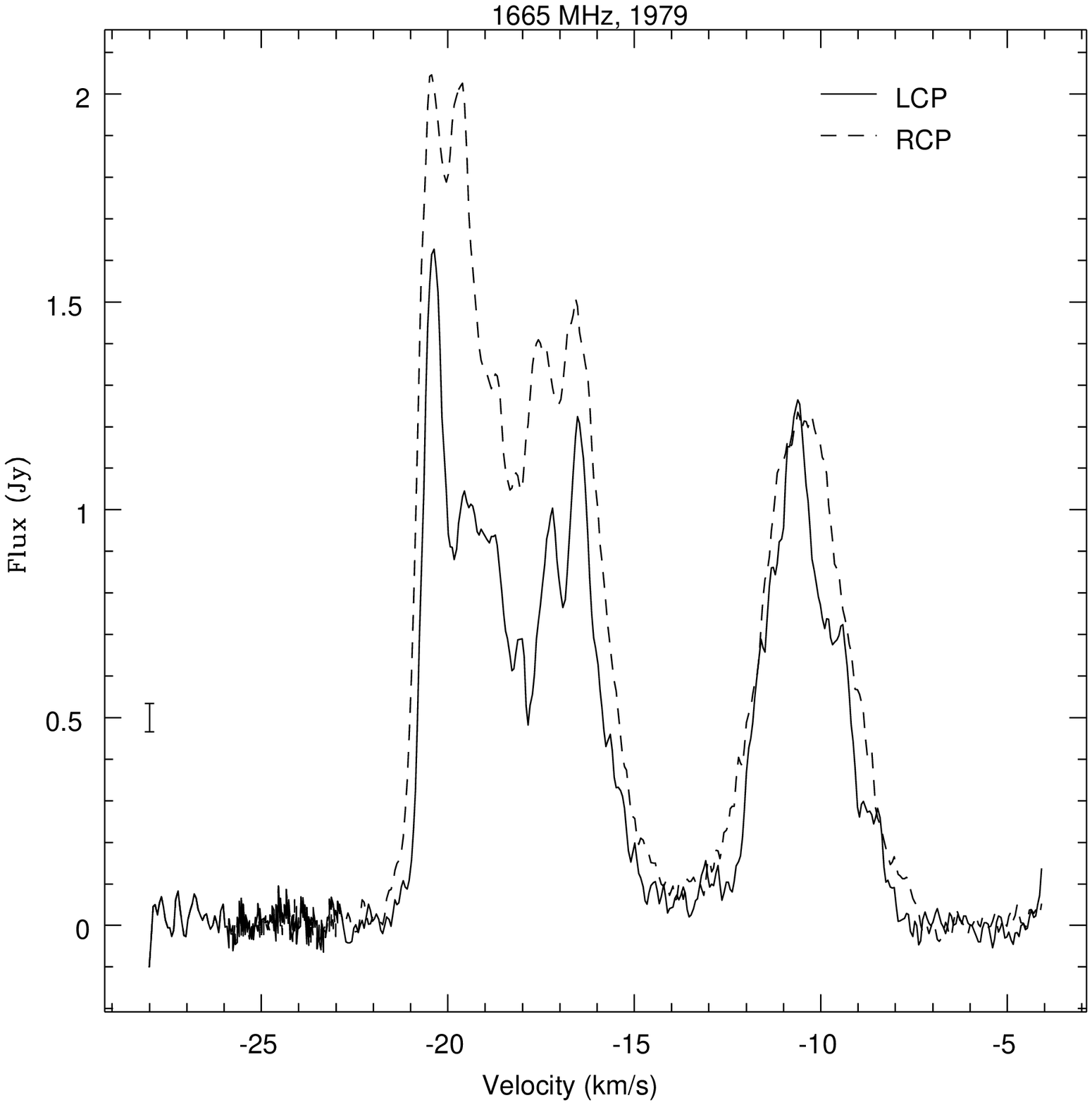 fig1d.ps fig1e.ps fig1f.ps fig1g.ps fig1h.ps fig1i.ps 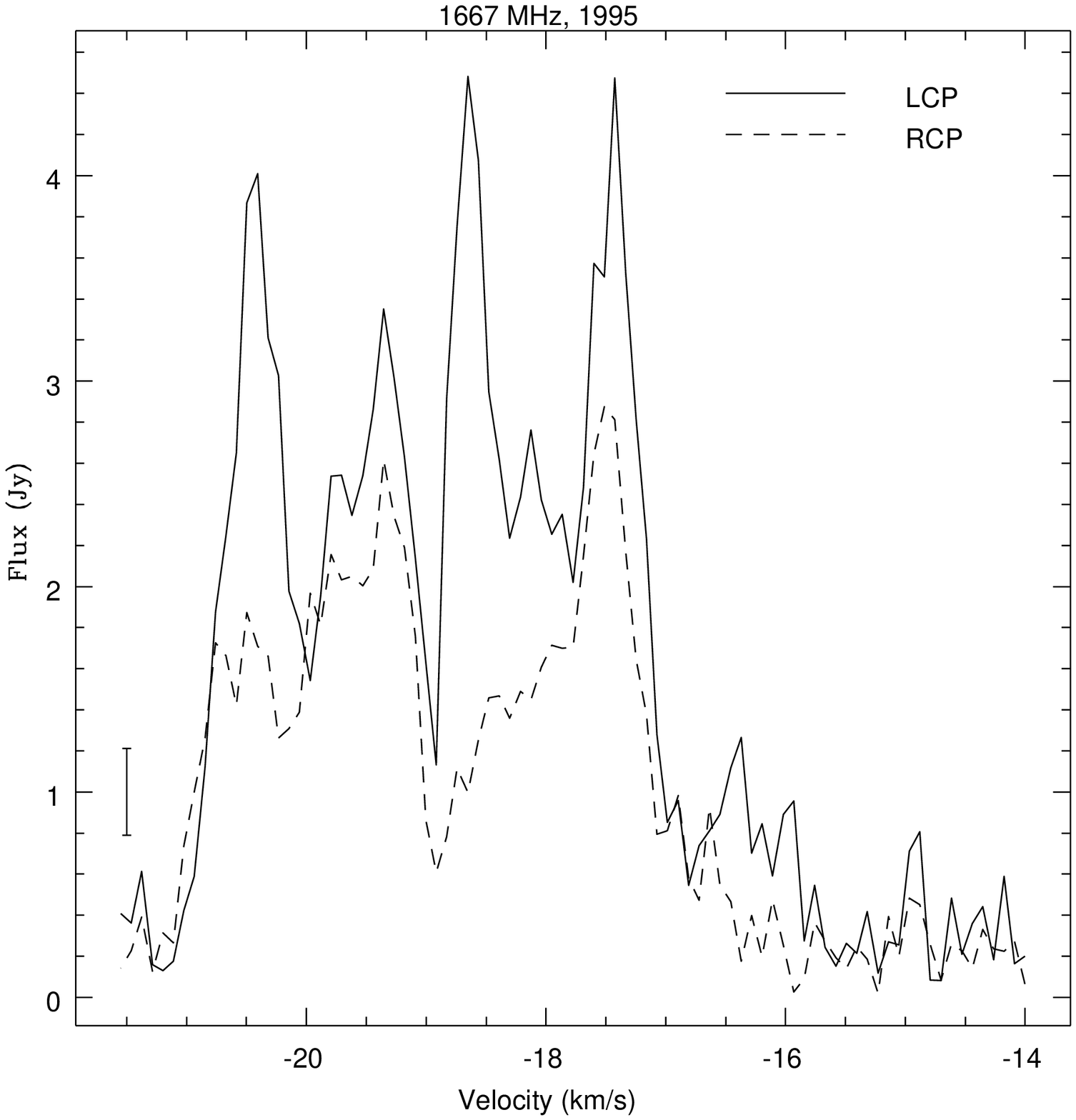] {The spectra of U Her at both polarizations, both IFs, and all epochs.  In all cases, the dashed line represents the right circularly polarized flux, while the solid line represents the left circularly polarized flux.  The rms error in each spectrum was calculated from a signal-free portion, and is represented by the error bar in the lower left of each plot.  At each epoch, some polarization is present, and it is apparent that spectral components can be individually polarized, since the polarization is not constant with velocity for each RCP-LCP pair of spectra.  This implies the presence of a magnetic field in the region, capable of Zeeman splitting the emission out of the line width.  For further discussion, see sections 4.3 and 5 in the text.  \label{spectra}}
\figcaption[fig2a.ps fig2b.ps] {Representative maps obtained with the VLBA.  Both maps are integrated across the spectral channels.  Figure 2a is the right polarized emission at 1665 MHz, and Figure 2b is the left circularly polarized emission at 1665 MHz. Contours are shown at 0.66, 1.98, 3.31, 4.63, and 5.95 Jy.  The center of both maps is located at approximately RA 16 25 47.477 and Dec 18 53 32.940 on the sky.  The peak flux in the maps is about 6.6 Jy/beam.\label{maps}}
\figcaption[fig3.eps] {Histogram of the results of the constant velocity model.  The horizontal axis shows the percentage of components which could be successfully projected forward in time to later epochs using a particular constant velocity.  The vertical axis shows the fractional number of trial velocities for which this percentage was obtained.  The results for both the simulated and the actual data are shown.  A reduced chi-square of 1.4 is found from these two binned distributions, implying that they are statistically identical.  Therefore, a constant velocity outflow can not be the complete explanation for the time variations of the OH masers in U Her. \label{histogram}}
\figcaption[fig4.eps] {The randomized velocity and magnetic fields.   \label{fields}}
\figcaption[fig5a.ps fig5b.ps fig5c.ps] {The parameter space of the line-of-sight plasma turbulence model.  The contours represent the filling factor for combinations of $C_{vel}$ and $C_{mag}/C_{vel}$.  Each plot represents a different value of the correlation parameter $\alpha$, and all plots shown have $K_0=25$.  Each contour indicates an increase of 0.1 in the filling factor, and the locations of the contours are approximate.  The shaded region indicates the values of $C_{vel}$ and $C_{mag}/C_{vel}$ for which the calculated filling factor matches the observed filling factor.  The size of this region decreases as the magnetic and velocity fields become more strongly correlated.   \label{turbmod}}

\end{document}